\documentclass[preprint,superscriptaddress]{revtex4}
\usepackage{graphicx}
\usepackage{array}
\usepackage{amssymb}
\usepackage{amsfonts}
\usepackage{amsmath}
\usepackage{mathrsfs}
\usepackage{color}
\usepackage{booktabs}
\usepackage{threeparttable}
\usepackage{multirow}
\usepackage{subfigure}
\usepackage{times}
\usepackage{epsfig}
\usepackage{threeparttable}
\usepackage{chngpage}
\usepackage{latexsym}
\usepackage{epstopdf}

\begin{document}

\title{Fundamental building blocks of controlling complex networks:
A universal controllability framework}

\author{Zhesi Shen}
\affiliation{School of Systems Science, Beijing Normal University,
Beijing, 100875, P. R. China}

\author{Wen-Xu Wang}\email{wenxuwang@bnu.edu.cn}
\affiliation{School of Systems Science, Beijing Normal University,
Beijing, 100875, P. R. China}
\affiliation{School of Electrical, Computer and Energy Engineering, Arizona
State University, Tempe, Arizona 85287, USA}
\affiliation{Business School, University of Shanghai for Science and Technology, Shanghai 200093, China}

\author{Chen Zhao}
\affiliation{School of Systems Science, Beijing Normal University,
Beijing, 100875, P. R. China}
\affiliation{College of Information Technology, Hebei Normal University, Hebei, 050024, P. R. China}

\author{Ying-Cheng Lai}
\affiliation{School of Electrical, Computer and Energy Engineering, Arizona
State University, Tempe, Arizona 85287, USA}
\affiliation{Department of Physics, Arizona State University,
Tempe, Arizona 85287, USA.}

\begin{abstract}
To understand the controllability of complex networks is a forefront problem relevant to different fields of science and engineering. Despite recent advances in network controllability theories, an outstanding issue is to understand the effect of network topology and nodal interactions on the controllability at the most fundamental level. Here we develop a universal framework based on local information only to unearth the most {\em fundamental building blocks} that determine the controllability. In particular, we introduce a network dissection process to fully unveil the origin of the role of individual nodes and links in control, giving rise to a criterion for the much needed strong structural controllability. We theoretically uncover various phase-transition phenomena associated with the role of nodes and links and strong structural controllability.
Applying our theory to a large number of empirical networks demonstrates that technological networks are more strongly structurally controllable (SSC) than many social and biological networks, and real world networks are generally much more SSC than their random counterparts with intrinsic resilience and adaptability as a result of human design and natural evolution.
\end{abstract}

\maketitle

Controlling complex networked systems has become a frontier field of
interdisciplinary research. The past few years have witnessed the
emergence of two theoretical frameworks: the structural controllability theory (SCT)~\cite{Lin:1974,SHIELDS_1976,LSB:2011} and the exact controllability
theory (ECT)~\cite{AM:book,Sontag_2013,YZDWL:2013}, both aiming to achieve full control
by identifying a minimum set of driver nodes to fully control the
dynamics of the underlying networked system. Indeed, the issue of
controllability has attracted a great deal of attention~\cite{NV:2012,
EMCCB:2012,YRLLL:2012,LSB:2012,WNLG:2012,RR:2012,LCZP:2012,SM:2013,
YZDWL:2013,JLCPSB:2013,JB:2013,PLSB:2013,LSB:2013,delpini2013evolution,jia2014connecting,
menichetti2014network,YZWDL:2014,Wuchty:2014,PZB:2014,posfai2014structural,
gao2014target,xiao2014edge,iudice2015structural,Yan_2015,zhao2015intrinsic,gu2015controllability,muldoon2016stimulation,liu2015control,basler2015control,skardal2015control,vinayagam2015controllability,wells2015control,whalen2015observability}.

Despite advances, there are fundamental issues that need to be
addressed to achieve full understanding of controlling
complex networks. First, at the most basic level, the effects of
nodes and links on control and their origins have not been fully understood, which
pertain to the problems of nodal and link classifications. To understand
the role of links in controllability is of particular interest, in
view of the recently developed notion of strongly structurally
controllability, i.e., to determine the circumstance under which the
network is strongly structurally controllable
(SSC)~\cite{hartung2012characterization,chapman2013strong,reissig2014strong}.
{ Especially, a network is said to be SSC if all link weights have no effect on the controllability and it is only determined by topology. A SSC network is generally more controllable and robust against the uncertainty and inaccessibility of link weights and external perturbation on links.}
However, a rigorous criterion has been missing
for determining if a network is SSC through controlling a minimum set
of driver nodes. Second, in spite of SCT and ECT, we lack a general
and efficient controllability framework, because SCT based on Lin's
classic controllability theory~\cite{Lin:1974} is applicable to
directed networks only and does not take into account the interaction
strengths along the links, while ECT based on the Popov-Belevitch-Hautus
rank condition~\cite{PBH} is sensitive to perturbation or computational
error (see Supplementary Note~1 for the details of SCT and ECT). In fact, the underlying connection between SCT and ECT has
remained unknown. These issues are fundamental to controlling
complex networks, but no existing theory can be adopted to address
them, calling for a drastically different approach to studying the
controllability of complex networks.

Here we develop an efficient controllability framework based on the
principle of structural dissection, which is universally applicable to
complex networks of arbitrary structures and link weights.
The framework goes much beyond SCT and
ECT by dissecting a network completely and revealing the role of nodes
and links in full control in a comprehensive but
straightforward manner. Specifically, the dissection process gives rise
to a hierarchy of configurations based solely on the local network
structure without requiring any global information. Central to the
process is the emergence of a core structure, in which the weights of
links play a determining role in controllability, whereas those of the links
outside the core have little effect on control. Thus, the hierarchical
structure classifies links into two categories in terms of SSC:
either essential or spareable. Mathematically, we prove that
a network consisting entirely of spareable links must be SSC, a generally sufficient
and necessary criterion for SSC. An interesting result is that, for
a SSC network, the SCT and ECT give exactly the same results, providing
a natural connection between the two well-established controllability
frameworks. Application of our dissection framework to real world networks
reveals that (i) technological networks are SSC, whereas many social
and biological networks are less so, and (ii) real world networks tend to
be much more SSC than their randomized counterparts. These results imply
that man-made networks intrinsically possess a higher degree of resilience
and adaptability, and natural evolution tends to contribute positively
to the emergence of SSC.

Additional results are the following. The hierarchical structure
classifies nodes into three categories in an extremely efficient manner:
critical, intermittent, and redundant in terms of their contribution to
controllability. Various phase transitions pertaining to the significance
of links and nodes have been uncovered. For both homogeneous and
heterogeneous networks, as a core emerges during the dissection process,
a discontinuous phase transition occurs. For random networks, the
phase transition point is universal but, for scale-free networks, it
depends on the degree distribution. For undirected random networks,
associated with nodal classification two phase transitions can arise.
A striking phenomenon is that, for different network sizes, the curves
characterizing the phase transitions intersect at a single point.
Based on the canonical control theory, we obtain a rigorous mathematical
proof for the dissection process. The analysis also enables us to make
theoretical predictions about the phase transitions associated with
full control.\\

\noindent
{\bf \large{Results}}\\
\noindent
{\bf Principle of structural dissection.}
The structural dissection process (SDP) dissects a network completely
to uncover a hierarchical structure that indicates the role of nodes and
links in the controllability of the network in an intuitive way.
The process consists of two orders (two stages) with respect to nodal
and link removal.

\begin{figure}[htbp]
\includegraphics[width=\linewidth]{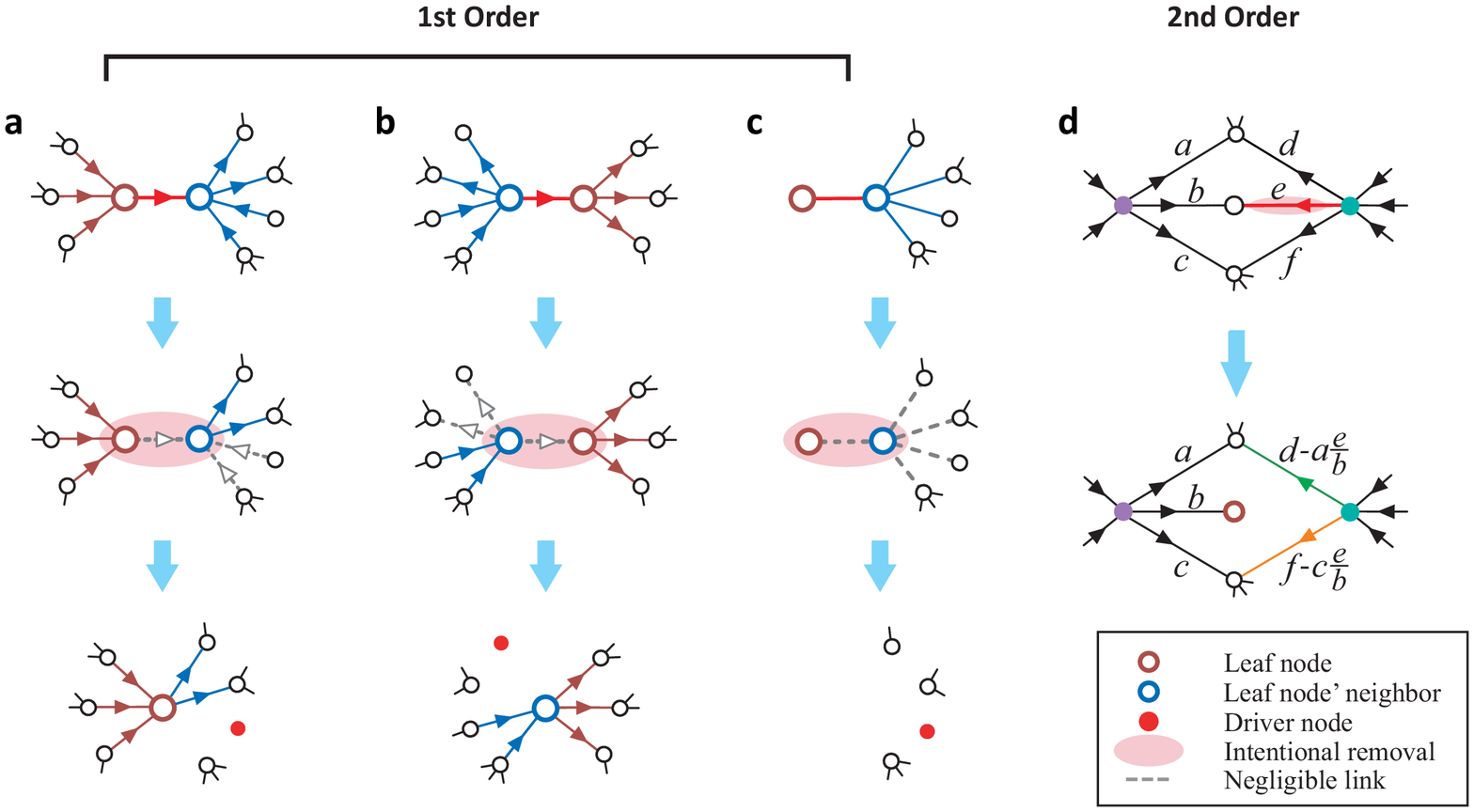}
\caption{{\bf Illustration of first- and second-order structural
dissection process.} First-order SDP for
({\bf a}) an out-leaf node, ({\bf b}) an in-leaf node, and ({\bf c}) a
leaf node in an undirected network. ({\bf d}) Second-order SDP, where the link with weight $e$
is removed, and the weights of the links ($d$ and $f$) are
updated. (See Supplementary Figs.~1 and~2 and Supplementary Note~2 for the corresponding adjacency
matrix representation and a mathematical proof of the SDP.) Driver
nodes and spareable links whose weights have no effect on
controllability can be identified. The configuration of the driver nodes
is not unique, but the configuration of the spareable links is.}
\label{fig:graph.transformation}
\end{figure}

For a directed network, the first-order SDP targets leaf
nodes and their associated links, as shown schematically in
Fig.~\ref{fig:graph.transformation}. For an out-leaf node, e.g., the dark
red node shown in Fig.~\ref{fig:graph.transformation}{\bf a} with unit
out-degree, we remove all the incoming links of the successor (the
blue node), and merge the out-leaf node (the dark red node) and its
successor (the blue node), and specify the merged node using the out-leaf
node's index. For an in-leaf node with unit in-degree, e.g., the dark red
node in Fig.~\ref{fig:graph.transformation}{\bf b}, we remove all the
outgoing links of the predecessor (the blue node), merge the in-leaf node (the
red node) and its predecessor (the blue one), and use the predecessor's
index to denote the resulting node. This procedure removes a single node
and some links (see Supplementary Note~2 and Supplementary Fig.~1 for first-order SDP in matrix representation).

For an undirected network, the first-order SDP entails removing the leaf
nodes and their neighbors iteratively from the network. For example, as
shown in Fig.~\ref{fig:graph.transformation}{\bf c}, a leaf node (the dark
red node) of degree one and its neighbors, together with all their links,
are removed. As a result, two nodes and some links are removed from the
network.

We repeat the first-order SDP until no leaf-nodes are left. For either
a directed or an undirected network, with respect to the end result of
the first-order SDP, two situations can arise. First, all remaining nodes
are isolated. In this case, the isolated nodes constitute a minimum set
of driver nodes and their number is $N_\text{D}$. We can prove
that the network is SSC in this case (see Supplementary Note~3 for proof of SSC).
Second, after the first-order SDP, in addition to the isolated nodes,
various cores without leaf-nodes can emerge. As a result, the isolated
nodes are the driver nodes, but nodes in the cores may not be. A
second-order SDP is then necessary to probe into the cores to identify
all the remaining driver nodes.

We can justify that the weights of the removed links by the first-order
SDP have no effect on the value of $N_\text{D}$. With respect to SSC,
it is reasonable to regard the removed links as "spareable".
Note that a second-order SDP is implemented only when cores are present.
Figure~\ref{fig:graph.transformation}{\bf d} schematically illustrates
the process, where the aim is to generate leaf nodes in the cores and this
can be accomplished by removing some links of the nodes with the smallest
degree. As shown in Fig.~\ref{fig:graph.transformation}{\bf d}, the link
marked as the pink circle is chosen to be removed, which induces a change
in the weight of each related link. The weights of the outgoing links of
the green node are updated. Specifically, the weights of the green and
orange links are changed to $d-a(e/b)$ and $f-c(e/b)$,
respectively (see Supplementary Fig.~2 and Supplementary Note~2 for second-order SDP in matrix
representation). Analogous to the first-order SDP, the second-order SDP
keeps $N_\text{D}$ and hence the controllability configuration invariant.
After leaf nodes arise, we can apply the first-order SDP continuously to
reduce the sizes of the cores until all cores disappear and all remaining
nodes are isolated.

The nodal and link removal processes take into account the link weights
in the remaining cores, which affect the value of $N_\text{D}$. For this
reason these links are regarded as "effective". If a network contains
effective links, it is not SSC.

Any network can be reduced to a set of isolated nodes through repeated and
alternative applications of the first- and second-order SDPs. In the end,
all the remaining isolated nodes are driver nodes, whose number is nothing
but $N_{\text D}$. This represents a general and efficient method to
identify a set of minimal driver nodes for arbitrary network structure
and link weights. Note that both the first- and second-order SDPs require
only information about the local network structure and are therefore highly
efficient. It is worth emphasizing that a SSC network contains only spareable
links, the weights of which have no effect on the value of $N_\text{D}$ and
the set of driver nodes. Thus, if a network is SSC, the SCT and ECT will
lead exactly to the same results of $N_\text{D}$ because of the null effect
of the link weights, establishing a natural connection between the two
existing controllability frameworks for complex networks.\\

\begin{figure}[htbp]
\includegraphics[width=0.8\linewidth]{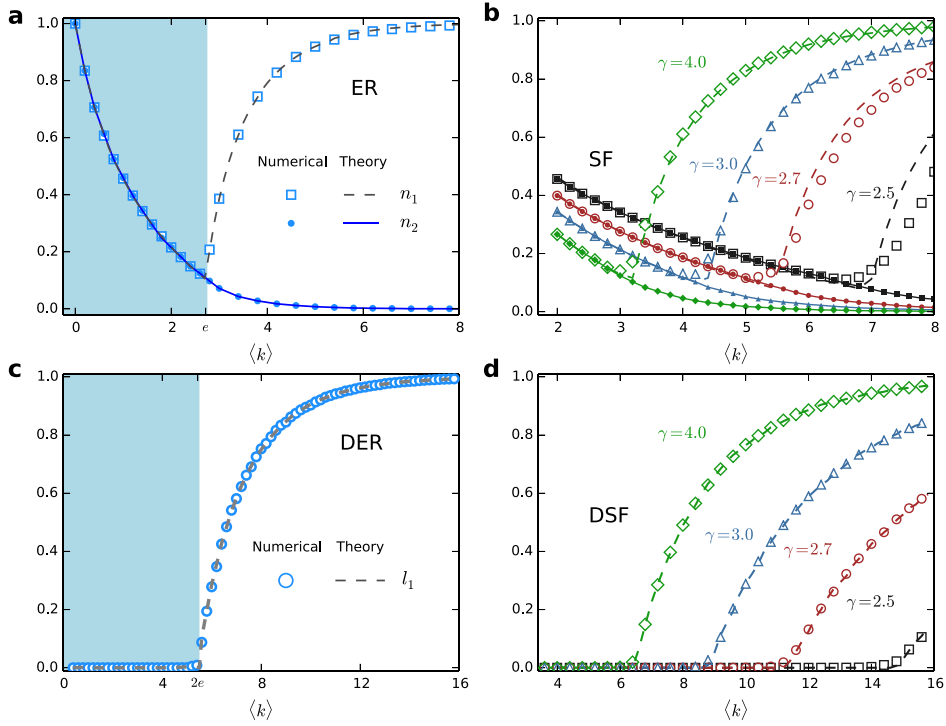}
\caption{{\bf Structural dissection of model networks.}
The fractions of the remaining nodes after first- and second-order
dissection processes, denoted as $n_1$ and $n_2$, respectively, versus
the average degree $\langle k \rangle$ for (a) an undirected ER random
network and (b) an undirected scale-free network. The fractions of the
remaining links ($l_1$) after the first-order dissection process versus
the average degree $\langle k \rangle$ are shown for (c) a directed random
network, and (d) a directed scale-free network. The unweighted random and
scale-free networks of size $N = 15000$ are generated from a static
model. The data points are result of averaging over 20 independent
realizations. The dashed curves are theoretical predictions from an
iterative equation based on the expected degree distribution. The symbols
are results from the network subject to structural dissection: open
circles and squares indicate the remaining fractions after the first-order
and second-order processes, respectively. The solid curves are results
obtained from ECT.}
\label{fig:core}
\end{figure}

\noindent
{\bf Controllability, link significance and SSC.}
We test SDP on both homogeneous and heterogeneous networks to discern
links with respect to SSC and identify driver nodes that determine controllability.
We first apply the first-order SDP and denote the fraction of the remaining
nodes and links as $n_1$ and $l_1$, respectively. Figure~\ref{fig:core}{\bf a}
shows, for Erd\"{o}s-R\'{e}nyi (ER) random networks, a phase-transition
in $n_1$ associated with the emergence of the cores at
$\langle k \rangle = e$. For $\langle k \rangle < e$, no core arises
after the first-order SDP terminates, indicating that the networks are
SSC. A core suddenly emerges at $\langle k \rangle = e$ and grows
gradually as $\langle k \rangle$ is increased. The vertical line
$\langle k \rangle = e$ thus separates two phases: (a) a phase on
the left side of the line, corresponding to SSC and the presence of
only spareable links that have no effect on the value of $n_\text{D}$,
and (b) a phase on the right side of the line with the emergence of
cores and effective links, violating the condition of SSC.
Figure~\ref{fig:core}{\bf c} shows the behavior of $l_1$ for directed
ER random networks, where a transition occurs at $\langle k \rangle = 2e$.
Figures~\ref{fig:core}{\bf b,d} show the results for undirected and
directed scale-free (SF) networks, respectively, where cases of different
exponents $\gamma$ associated with the power-law degree distribution were
considered. The phenomenon of phase transition between SSC and non-SSC
persists in all the cases but with a difference from the ER networks: the
critical transition point now depends on the value of the degree
distribution exponent $\gamma$.

\begin{figure}
\includegraphics[width=\linewidth]{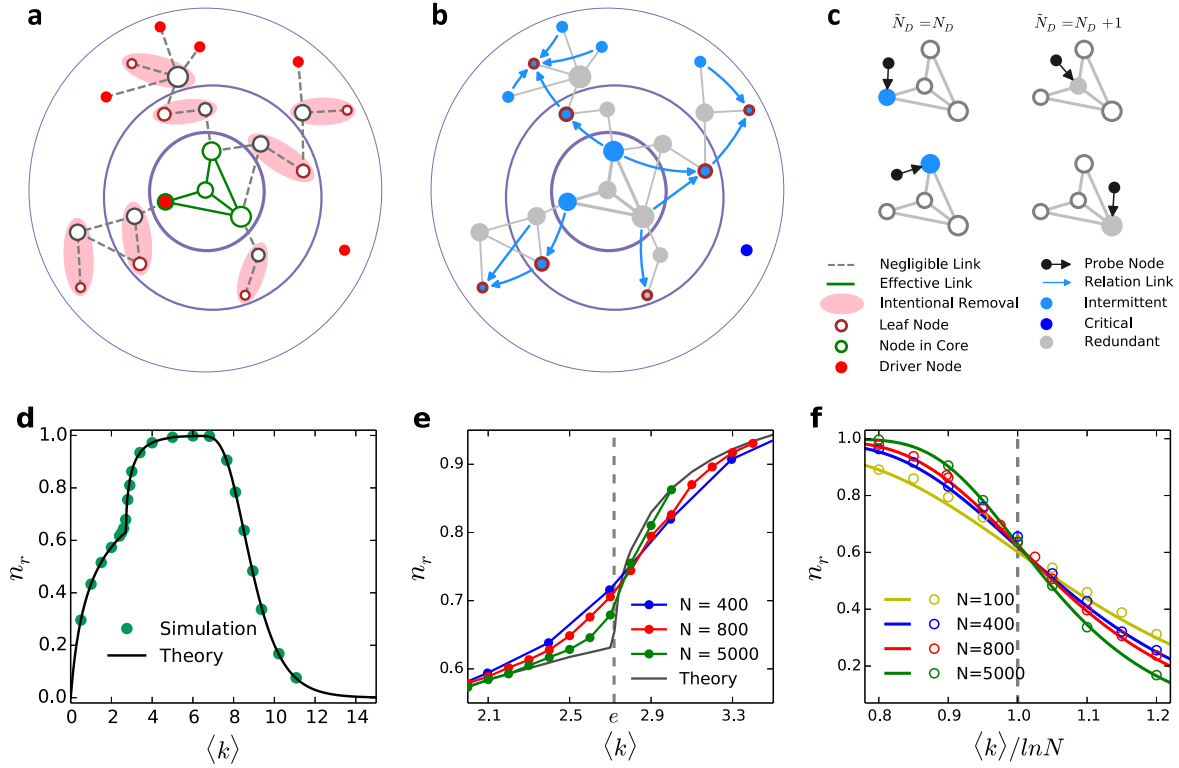}
\caption{{\bf Structural dissection, hierarchical relation and nodal
classification.} ({\bf a}) First- and second-order structural dissection
of an undirected network. A minimum set of driver nodes can be identified.
({\bf b}) The hierarchical relation network that specifies the node
categories, which consists of directed links pointing to
the removed leaves from their neighbors' neighbors. The category of a
node is determined by its upstream neighbors in the relation network. A
node is intermittent if and only if at least one of its upstream
neighbors is an intermittent node. If all the upstream neighbors of a node
in the relation network are redundant, the node itself is redundant.
({\bf c}) Determination of node categories in the core through the addition
of an external probing node. After adding a probing node pointing at a
node, if the number of the driver nodes in the core is increased by 1, the
node pointed at is redundant. Otherwise, if the number of driver nodes
does not change, the node is classified as intermittent.
Node classification for ER random networks: ({\bf d}) The fraction $n_{\rm r}$ of redundant nodes as a function of the average degree $\langle k \rangle$ for undirected ER random networks, ({\bf e}) the phase transition at $\langle k \rangle=e$ and ({\bf f}) the phase transition at $\langle k \rangle/\text{ln}N$. At the
phase-transition point, the curves of $n_{\rm r}$ versus $\langle k \rangle$
for different network sizes cross each other at exactly the same point, unequivocal indication of a phase transition. The data points are result of averaging over 500 independent network realizations.
Curves are theoretical prediction based on the expected degree distribution in the $N\rightarrow \infty$
limit (see Supplementary Note~2 for analytical expression), where the decreasing part is based on the degree distribution for $N=15000$.}
\label{fig:hierarchy}
\end{figure}

To obtain a clearer picture of the SDPs, we illustrate in
Fig.~\ref{fig:hierarchy}{\bf a} the first-order SDP, where we see that
a hierarchical structure arises and a core emerges in the central area.
During the process, partial driver nodes and all effective links can be
distinguished, where the former is the surviving isolated nodes and the
latter belong to the core. Repeated and alternative applications of the
first- and second-order SDPs for both homogeneous and heterogeneous
networks yield all driver nodes for full control. The cores vanish
and all the remaining nodes are isolated, whose number is $N_\text{D}$,
as exemplified in Figs.~\ref{fig:core}{\bf a}-{\bf d}, which agrees exactly
with the results from ECT but in a highly efficient way. We were able to
obtain theoretical predictions for the phase transitions and the core
sizes, which agree well with the numerical results for all cases
(see Supplementary Figs.~3 and~4 and Supplemenary Note~2).\\

\noindent
{\bf Nodal significance.}
The significance of the nodes can be measured in terms of their role in
controllability. A previous work~\cite{JLCPSB:2013} suggested three types
of nodes: critical, redundant and intermittent (the detailed definitions
can be found in Supplementary Note~4). A brute-force search requiring
global information was introduced for node classification for directed
networks. However, the method is not applicable to undirected networks.
The emerging hierarchical structure as a result of applying our first-order
SDP opens up a new way to classify nodes in networks of arbitrary structure
in an extremely efficient manner. Specifically, SDP allows us to
establish a {\em directed relation network} among nodes based on local
information only, where each removed leaf is pointed at by its neighbors'
neighbors, and the category of the nodes is determined by their
predecessors in the relation network. The remaining isolated nodes after
first-order SDP are intermittent nodes. For the nodes removed during the
SDP, their categories can also be determined. In particular, for an
uncategorized node, if there is at least one intermittent node in its
predecessors in the relation network, it will be classified an intermittent
node. We repeat this procedure until no more intermittent nodes are
categorized, and then the remaining nodes are redundant.

Figure~\ref{fig:hierarchy}({\bf b}) illustrates the relation network
built from the outside to insider layers based on the principle
illustrated in Fig.~\ref{fig:hierarchy}({\bf a}). The directions of the
relation links are all from the leaves' neighbors' neighbors to the
removed leaves, and from the insider to the outside layers. Note that
the relation network is not unique, but any of its configurations can
place each node into a unique category (see Supplementary Note~4 for
how to build the relation network and the corresponding mathematical
proof). The categories of all nodes are determined by the remaining
isolated nodes and nodes in the core through the relation network. The
categories of the nodes in the core can be determined by adding a probing
node pointing to it (See Supplementary Figs.~5 and~6). As shown in Fig.~\ref{fig:hierarchy}({\bf c}),
if the probing node is added to a redundant node, we will have
$N_{\rm D} \rightarrow N_{\rm D}+1$ for the core. But if the probing
node is added to an intermittent node, $N_{\rm D}$ will remain unchanged.
In general, the core is much smaller than the whole network, leading to
a much higher efficiency than with the brute-force search.

We focus on nodal classification in undirected networks, a feat that has
not been achieved prior to our work. We find some nontrivial phenomena in
ER networks, as demonstrated in Figs.~\ref{fig:hierarchy} ({\bf d}) - ({\bf f}), where $n_{\rm r}$,
the fraction of redundant nodes, versus the average degree is shown for
different values of the network size. We see that $n_{\rm r}$ increases
with the average degree $\langle k \rangle$ when it is small but decreases
when $\langle k \rangle$ exceeds a threshold. Two distinct phase
transitions can be identified, which occur at $\langle k\rangle =e$ and
$\langle k \rangle/\ln{N}=1$, respectively.
Fig.~\ref{fig:hierarchy}({\bf e}) shows the behavior of $n_{\rm r}$ for
$\langle k \rangle=e$, where we see that all curves for different network
size $N$ cross each other at the same point - the defining signature of
phase transition. If we normalize the average degree $\langle k \rangle$
by $\text{ln} N$, the curves associated with different system sizes cross
at $\langle k \rangle/\text{ln}N=1$, as shown in Fig.~\ref{fig:hierarchy}({\bf f}),
indicating another phase transition at this critical point. We offer
a theoretical explanation for the behavior of $n_{\rm r}$ versus
$\langle k \rangle$, which agree well with the numerical results
(see Supplementary Note~4 for the theoretical expression of $n_{\rm r}$).\\

\begin{figure}
\includegraphics[width=\linewidth]{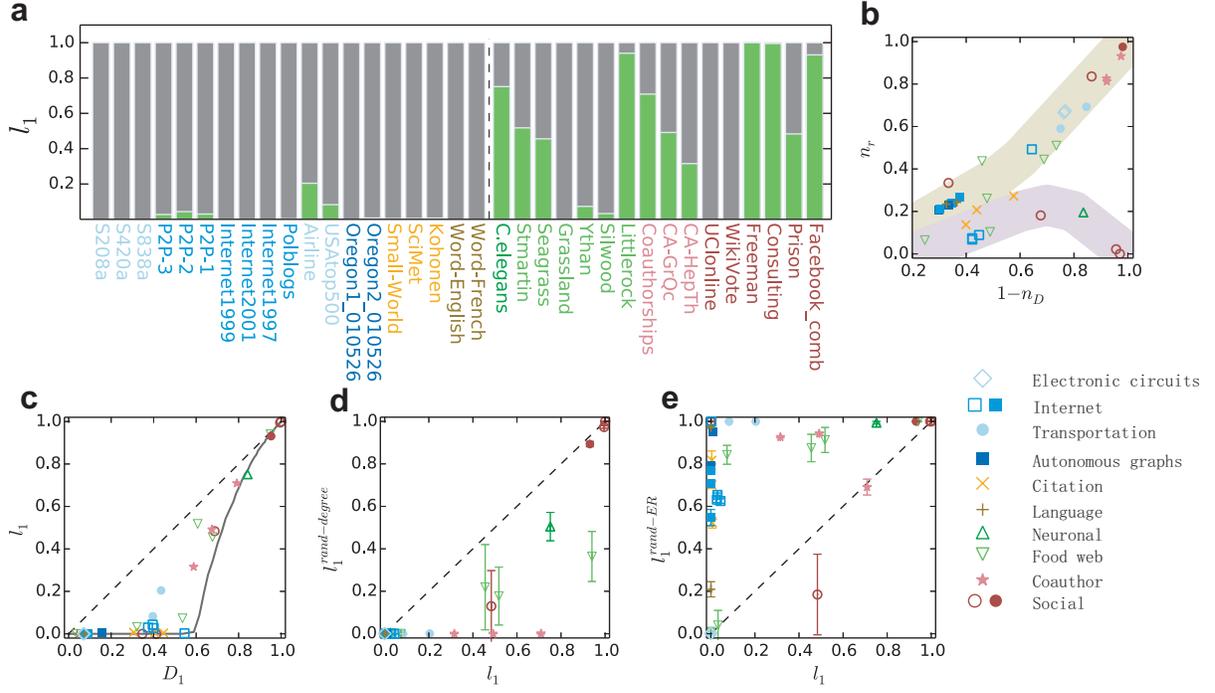}
\caption{{\bf Structural dissection analysis of real-world networks.}
({\bf a}) The fractions of remaining links ($l_1$) after the first-order
SDP for a number of real-world networks, where SSC can be characterized
by the value of $l_1$. ({\bf b}) Fraction $n_{\rm r}$ of redundant nodes
versus $1-n_{\rm D}$. ({\bf c}) The relation between $l_1$ and the
fraction $D_1$ of remaining links after the first round of the 1st-order
SDP ($D_1$). ({\bf d}) Fraction of remaining links after 1st-order SDP,
$l^{\rm rand-degree}_1$, obtained from the degree-preserving randomized
network in comparison with the values of $l_1$. ({\bf e}) The fraction
of remaining links after the first-order SDP, $l^{\rm rand-ER}_1$,
obtained from fully randomized version in comparison with the values of
$l_1$.}
\label{fig:real}
\end{figure}

\noindent
{\bf Application of SDP to real networks.}
Table~\ref{tab:real} and Fig.~\ref{fig:real} show the results from applying SDP to a variety of real-world networks, directed or undirected, where $n_2$ and $n_2'$
are the fractions of the remaining nodes after the SDP for identical
link weights (equivalent to unweighted networks) and random link weights,
respectively (Details about these networks are included in
Supplementary Table~{\bf 1} and Note~ 5). We see that, for all the networks, the
values of $n_2$ and $n_\text{D}$ from the ECT~\cite{YZDWL:2013} are
exactly equal to each other. For directed networks with random weights,
the values of $n_2'$ are exactly the same as those of $n_\text{D}$
from the SCT~\cite{LSB:2011,YZDWL:2013}. The consistency in these
results validates our SDP method for characterizing controllability.

In Table~\ref{tab:real}, $l_1$ is a key index for SSC. In particular, if a network has $l_1=0$, then all the links are sparable and the
network is SSC. In this case, we have $n_1=n_2$ because of the absence of cores in the real networks and $n_2=n_2'$ because of the null
effect of the link weights. From the values of $l_1$ in
Fig.~\ref{fig:real}({\bf a}), we find that almost all technological
networks are SSC with their $l_1$ values close to zero. In contrast,
most social and biological networks are not SSC because their values of
$l_1$ are not close to zero. These results are consistent with the
intuition that the man-made networks, due to their design optimization,
are generally more SSC than natural and self-organized networks.
In addition, as shown in Table~\ref{tab:real}, non-SSC networks ($l_1 >0$) with random link weights are always more controllable than unweighted non-SSC networks, i.e., $n_2' < n_2$. This is because of the elimination of linear correlations by random weights comparing with identical weights.

Performing node classification for the real world networks, we find that
the number of critical nodes tends to be much smaller than
that of intermittent nodes (see $n_{\rm c}$, $n_{\rm i}$ and $n_{\rm r}$ in Table~\ref{tab:real}). Except for a few social networks, the fraction of intermittent nodes ($n_{\rm i}$) is much larger than that of critical nodes ($n_{\rm c}$), indicating that in general there are many configurations of the driver nodes to
realize full control. That is, there is flexibility in controlling
real world networks, in spite of the existence of a fraction of
inaccessible nodes on which external input signals cannot be imposed.
Another result is that, for undirected networks, the quantity $n_{\rm D}$
is negatively correlated with $n_{\rm r}$ [the solid symbols in the
upper branch in Fig.~\ref{fig:real}({\bf b})], whereas for directed
networks, there exists a bifurcation in the values of $n_{\rm D}$ from
those of $n_{\rm r}$. There is thus a generic difference between
directed and undirected networks. Figure~\ref{fig:real}({\bf c}) shows
the relation between the fraction $D_1$ of the remaining links after
the first round of the first-order SDP and $l_1$ obtained from all
rounds of the process. The results indicate that the quantity $D_1$
plays a dominant role in generating SSC
as characterized by $l_1$. More specifically, when $D_1$ is less than
a critical value (about 0.4), the values of $l_1$ are quite close to
zero, indicating that the network is SSC. In contrast, when $D_1$ exceeds
a critical value, $l_1$ increases rapidly and the network is not SSC.

Figure~\ref{fig:real}({\bf d}) shows that the real networks become
more SSC when the nodal degrees are fixed but the links are randomized.
The reason can be attributed to the existence of the core structures
but link randomization among the nodes would
destroy the cores so as to enhance the degree of SSC. However, if the
nodal degrees are allowed to vary, the result would be drastically
different in that the networks can be much more SSC than their
completely randomized counterparts, as shown in Fig.~\ref{fig:real}({\bf e}).
A plausible reason is that the networks, regardless of whether they
are formed through self-organization (e.g., social networks), natural
evolution (e.g., biological networks), or intentional design (e.g.,
technological networks), all have the tendency to be SSC. It is worth
noting that, for SSC, degree distribution does not play a significant
role [Fig.~\ref{fig:real}({\bf d})], which is characteristically
different from the conventional structural controllability. Instead,
$D_1$ dominates $l_1$ [Fig.~\ref{fig:real}({\bf c})], which can be
explained in terms of the first-order SDP (the results for the randomized networks can be found in Supplementary Fig.~7 and Supplementary Note~5). \\

\noindent
{\large\bf Discussion}\\
Our structural dissection paradigm provides an effective way to identify
the fundamental building blocks in complex networks of arbitrary topology
and weights, which determine the network controllability. Comparing with
the existing SCT and ECT frameworks, our approach offers a much deeper
understanding of control of complex networks. Our study indicates that
local structures are sufficient for precisely pinning down the key
nodes and links for determining controllability, in contrast to SCT and ECT that
require a global search and optimization scheme for nodal classification
with respect to controllability. Particularly, based on the local information,
structural dissection can be done in an extremely efficient manner, which
gives rise to a hierarchical structure that classifies links and nodes
into different categories with respect to their role in controllability. In general, two categories of links arise, where the weights of
the effective links play a role in controllability but those of the spareable
links do not. A network is SSC if it contains only spareable links, a
sufficient and necessary criterion for determining SSC efficiently. The criterion
naturally establishes an intrinsic connection between SCT and ECT in the
sense that, for SSC networks, they give rise to the same controllability
result.

Some striking results are obtained when we apply our SSC criterion to
real-world networks. In particular, technological networks tend
to be more SSC, whereas most social networks and biological networks are less.
This indicates that man-made networks, due to the underlying design
process to optimize performance, are generally more controllable than
networks formed through natural evolution or self-organization. Our structural dissection process
can also classify nodes efficiently into three categories { in the
underlying hierarchical relation structure}, regardless of whether the network is
directed or undirected. { Two phase transitions associated with the redundant
nodes in undirected random networks are uncovered, and real directed and undirected networks are drastically different with respect redundant nodes.}



There are some outstanding questions that need to be addressed to further
advance the field of controlling complex networks. First and foremost,
finding a framework to generalize the linear control theory to nonlinear
dynamical networks is urgent. For a nonlinear dynamical system, the
classical canonical control theory is not applicable. It is thus not
clear whether our structural dissection framework can be applied to arbitrary
nonlinear networks. Second, in existing works, controllability is referred
to as the implementation of control through a minimum set of driver
nodes. However, there are other definitions for controllability such as
one in terms of the control energy. The trade-off
between a smaller number of driver nodes and high energy cost provides
a physically more reasonable way to define and realize better control.
Third, for very dense networks where zeros no longer dominate the eigenvalue spectrum,
knowledge about the eigenvalues is needed to implement the structural
dissection process (see Supplementary Note~6 and Supplementary Figs.~8 and~9 for SDP in situations where zero is not the dominant eigenvalue), calling for an
improved approach independent of the eigenvalues. In spite of the open
issues, our universal controllability framework goes much beyond the
existent ones and provides a deeper understanding of the controllability of
complex networks at the most fundamental level.\\


\bibliographystyle{naturemag}

\noindent
{\bf \large Acknowledgement}\\
We thank Dr. Y.-Y. Liu for valuable discussion.
This work was supported by NSFC under Grant No.~11105011 and CNNSF under
Grant No.~61074116. Y.-C.L. was supported by ARO under Grant
No.~W911NF-14-1-0504.\\

\noindent
{\bf \large{Author contribution}}\\
W.-X.W. and Z.S. contribute equally.
W.-X.W., Z.S., C.Z., and Y.-C.L. designed research;
Z.S. performed computations. Z.S., C.Z., and W.-X.W. did
analytic derivations and analyzed data.
Z.S., W.-X.W., and Y.-C.L. wrote the paper.\\

\noindent
{\bf \large{Additional information}}\\
\noindent
{\bf Supplementary Information} accompanies this paper as http://www.nature.com/
naturecommunications\\

\noindent
{\bf Competing financial interests:} The authors declare no competing financial interests.\\

\noindent
{\bf reprints and permission} information is available online at http://npg.nature.com/
reprintsandpermissions/\\

\noindent
{\bf How to cite this article:}

\newpage

\begin{table}[H] \scriptsize
\linespread{0.9}
\centering{}
\caption{\small {\bf Structural dissection of real world networks.}
For each network, its type and name, the number of nodes ($N$), and the
number of links ($L$) are shown. The quantities $n_1$ and $l_1$ are the
fractions of the remaining nodes and edges after first-order
dissection, respectively. The quantities $n_2$  and $n_2'$ are the
fractions of the remaining nodes after second-order dissection for
networks with uni-weights and random weights, respectively. The
quantities $n_{\text{c}}$, $n_{\text{i}}$ and $n_{\text{r}}$ are the fractions of the critical,
intermittent and redundant nodes, respectively. For data sources and
references, see Supplementary Table 1 and Supplementary Note 5.}
\begin{center}
\resizebox{\columnwidth}{!}{
\begin{tabular}{cccccccccccc} 
\hline
\hline
& Name & Type & $N$ & $L$ & $n_1$ & $l_1$ & $n_2$ & $n_2'$ & $n_{\text{c}}$ & $n_{\text{i}}$ & $n_{\text{r}}$\\
\hline
\multirow{3}{*}{Electronic circuits} 	&	 S208a 	&	 D 	&	122	&	189	&	0.2377	&	{\textcolor{blue}{	0	}}	&	0.2377	&	0.2377	&	0.082	&	 0.2541 & 0.6639\\
	&	 S420a 	&	 D 	&	252	&	399	&	0.2341	&	{\textcolor{blue}{	0	}}	&	0.2341	&	0.2341	&	0.0714	&	 0.2579 & 0.6706\\
	&	 S838a 	&	 D 	&	512	&	819	&	0.2324	&	{\textcolor{blue}{	0	}}	&	0.2324	&	0.2324	&	0.0664	&	 0.2598 & 0.6738\\
\hline																						
\multirow{2}{*}{Autonomous graphs} 	&	 Oregon1-010526 	&	 U 	&	11174	&	23409	&	0.7036	&	{\textcolor{blue}{	0.0009	}}	&	0.7028	&	0.7027	&	0	&	 0.7928 & 0.2072\\
	&	 Oregon2-010526 	&	 U 	&	11461	&	32730	&	0.6714	&	{\textcolor{blue}{	0.0074	}}	&	0.6661	&	0.6659	&	0	&	 0.7715 & 0.2285\\
\hline																						
\multirow{7}{*}{Internet}	&	 P2P-3 	&	 D 	&	8717	&	31525	&	0.5845	&	{\textcolor{blue}{	0.0273	}}	&	0.5778	&	0.5774	&	0.0091	&	 0.9235 & 0.0675\\
	&	 P2P-2 	&	 D 	&	8846	&	31839	&	0.5861	&	{\textcolor{blue}{	0.0435	}}	&	0.5779	&	0.5778	&	0.0133	&	 0.9131 & 0.0736\\
	&	 P2P-1 	&	 D 	&	10876	&	39994	&	0.5645	&	{\textcolor{blue}{	0.0298	}}	&	0.5531	&	0.552	&	0.0018	&	 0.9092 & 0.0889\\
	&	 Internet1997 	&	 U 	&	3015	&	5156	&	0.6245	&	{\textcolor{blue}{	0	}}	&	0.6245	&	0.6245	&	0	&	 0.7333 & 0.2667\\
	&	 Internet1999 	&	 U 	&	5357	&	10328	&	0.6532	&	{\textcolor{blue}{	0.001	}}	&	0.6517	&	0.6517	&	0	&	 0.7616 & 0.2384\\
	&	 Internet2001 	&	 U 	&	10515	&	21455	&	0.6999	&	{\textcolor{blue}{	0.001	}}	&	0.6988	&	0.6987	&	0	&	 0.7874 & 0.2126\\
	&	 Polblogs 	&	 D 	&	1224	&	19025	&	0.3668	&	{\textcolor{blue}{	0.0024	}}	&	0.3595	&	0.3562	&	0.1912	&	 0.3162 & 0.4926\\
\hline																						
\multirow{2}{*}{Transportation} 	&	 USAtop500 	&	 U 	&	500	&	2980	&	0.508	&	{\textcolor{blue}{	0.0829	}}	&	0.264	&	0.25	&	0	&	 0.41 & 0.59\\
	&	 Airline 	&	 U 	&	332	&	2126	&	0.5181	&	{\textcolor{blue}{	0.2041	}}	&	0.1747	&	0.1536	&	0	&	 0.3072 & 0.6928\\
\hline																						
\multirow{3}{*}{citation} 	&	 Small-World 	&	 D 	&	233	&	994	&	0.6094	&	{\textcolor{blue}{	0.004	}}	&	0.6052	&	0.6009	&	0.0086	&	 0.8541 & 0.1373\\
	&	 SciMet 	&	 D 	&	2729	&	10413	&	0.4306	&	{\textcolor{blue}{	0.0046	}}	&	0.4251	&	0.4236	&	0.1261	&	 0.6017 & 0.2723\\
	&	 Kohonen 	&	 D 	&	3772	&	12731	&	0.5673	&	{\textcolor{blue}{	0.0068	}}	&	0.562	&	0.5604	&	0.0673	&	 0.7256 & 0.2071\\
\hline																						
\multirow{2}{*}{language} 	&	 Word-English 	&	 D 	&	7381	&	46281	&	0.6345	&	{\textcolor{blue}{	0	}}	&	0.6345	&	0.6345	&	0.0167	&	 0.7404 & 0.2429\\
	&	 Word-French 	&	 D 	&	8325	&	24295	&	0.6739	&	{\textcolor{blue}{	0.0003	}}	&	0.6736	&	0.6734	&	0.0432	&	 0.7223 & 0.2345 \\
\hline
\cline{1-12}
\multirow{6}{*}{Food Web} 	&	 grassland 	&	 D 	&	88	&	137	&	0.5227	&	{\textcolor{blue}{	0	}}	&	0.5227	&	0.5227	&	0.0114	&	 0.7273 & 0.2614\\
	&	 ythan 	&	 D 	&	135	&	601	&	0.5926	&	{\textcolor{blue}{	0.0732	}}	&	0.5185	&	0.5111	&	0.0074	&	 0.8889 & 0.1037\\
	&	 silwood 	&	 D 	&	154	&	370	&	0.7727	&	{\textcolor{blue}{	0.0324	}}	&	0.7662	&	0.7532	&	0.0065	&	 0.9286 & 0.0649\\
	&	 stmartin 	&	 D 	&	45	&	224	&	0.7111	&	{\textcolor{blue}{	0.5179	}}	&	0.4	&	0.3111	&	0.0222	&	 0.5333 & 0.4444\\
	&	 seagrass 	&	 D 	&	49	&	226	&	0.6939	&	{\textcolor{blue}{	0.4558	}}	&	0.3265	&	0.2653	&	0.0204	&	 0.4694 & 0.5102\\
	&	 littlerock 	&	 D 	&	183	&	2494	&	0.9781	&	{\textcolor{blue}{	0.9403	}}	&	0.7541	&	0.541	&	0.0055	&	 0.5574 & 0.4372\\
\hline																						
neuronal 	&	 C.elegans Neuronal 	&	 D 	&	297	&	2345	&	0.8485	&	{\textcolor{blue}{	0.7514	}}	&	0.165	&	0.165	&	0.0909	&	 0.7138 & 0.1953\\
\hline																						
\multirow{3}{*}{coauthor} 	&	 CA-HepTh 	&	 U 	&	9877	&	25998	&	0.4459	&	{\textcolor{blue}{	0.3156	}}	&	0.1314	&	0.0778	&	0	&	 0.1874 & 0.8126\\
	&	 CA-GrQc 	&	 U 	&	5242	&	14496	&	0.4914	&	{\textcolor{blue}{	0.4918	}}	&	0.2224	&	0.0792	&	0	&	 0.174 & 0.8260\\
	&	 coauthorships 	&	 U 	&	1461	&	2742	&	0.6578	&	{\textcolor{blue}{	0.709	}}	&	0.4613	&	0.0253	&	0	&	 0.0684 & 0.9316 \\
\hline																						
\multirow{6}{*}{social} 	&	 facebook-combined 	&	 U 	&	4039	&	88234	&	0.9445	&	{\textcolor{blue}{	0.931	}}	&	0.0255	&	0.0191	&	0	&	 0.0248 & 0.9752\\
	&	 Freeman-1 	&	 D 	&	34	&	695	&	1	&	{\textcolor{blue}{	1	}}	&	0.0294	&	0.0294	&	0	&	 1 & 0\\
	&	 consulting 	&	 D 	&	46	&	879	&	0.9783	&	{\textcolor{blue}{	0.9954	}}	&	0.0435	&	0.0435	&	0	&	 0.9783 & 0.0217\\
	&	 UCIonline 	&	 D 	&	1899	&	20296	&	0.3233	&	{\textcolor{blue}{	0	}}	&	0.3233	&	0.3233	&	0.0195	&	 0.7994 & 0.1811\\
	&	 prison 	&	 D 	&	67	&	182	&	0.5224	&	{\textcolor{blue}{	0.4835	}}	&	0.1343	&	0.1343	&	0.1045	&	 0.0597 & 0.8358\\
	&	 WikiVote 	&	 D 	&	7115	&	103689	&	0.6656	&	{\textcolor{blue}{	0	}}	&	0.6656	&	0.6656	&	0.6654	&	 0.0006 & 0.3341\\
\hline
\end{tabular}
}
\end{center}
\label{tab:real}
\end{table}

\end{document}